# On-chip erbium-doped lithium niobate microdisk lasers

Qiang Luo [1], ZhenZhong Hao [1], Chen Yang [1], Ru Zhang [1], DaHuai Zheng [1], ShiGuo Liu [1], HongDe Liu [1], Fang Bo [1,2,3,*], YongFa Kong [1,*], GuoQuan Zhang [1,*] And JingJun Xu [1,*]

[1] *MOE Key Laboratory of Weak-Light Nonlinear Photonics, TEDA Institute of Applied Physics and School of Physics, Nankai University, Tianjin 300457, China;*
[2] *Collaborative Innovation Center of Extreme Optics, Shanxi University, Taiyuan 030006, China*
[3] *Collaborative Innovation Center of Light Manipulations and Applications, Shandong Normal University, Jinan 250358, China*

Erbium-doped lithium niobate high-$Q$ microdisk cavities were fabricated in batches by UV exposure, inductively coupled plasma reactive ion etching and chemo-mechanical polishing. The stimulated emission at 1531.6 nm was observed under the pump of a narrow-band laser working at 974 nm in erbium-doped lithium niobate microdisk cavity with threshold down to 400 µW and a conversion efficiency of $3.1\times 10^{-4}$ %, laying the foundation for the LNOI integrated light source research.

**Lithium niobite, LNOI, microcavities, laser**

## 1 Introduction

As an excellent optical crystal material, lithium niobate (LN) has the advantages of small absorption coefficient (0.02 cm$^{-1}$ at 1064 nm), wide transparent window (0.35-5 µm), high nonlinear coefficient ($d_{33}$ = -41.7 pm/V) and good electro-optic ($r_{33}$ = 31.9 pm/V) [1], acousto-optic, photorefractive effects, etc. Benefiting from the commercial production of LN on insulator (LNOI), the researches on LNOI integrated optical devices have increased explosively. For example, Lin et. al fabricated the first LNOI microdisk cavity using femtosecond laser micromachining followed by focused ion beam milling [2]. Subsequently, an inductively coupled plasma reactive ion etching (ICP-RIE) process was introduced to prepare LNOI microdisk cavity [3], and combined with photolithography to achieve batch production [4]. Recently, with the assistance of chemo-mechanical polishing (CMP), the quality ($Q$) factors of LNOI microdisk cavities have been improved up to $10^7$ [5,6]. Using the nonlinearity of LN, various nonlinear optical effects including sum frequency generation, second harmonic generation and difference frequency generation, four-wave mixing are realized in LNOI microdisks, microrings and waveguides [7-16]. Integrated LN electro-optic modulators with high operating frequency with CMOS-compatible voltages were achieved [17]. Subsequently, Cai et al. designed an MZI electro-optic modulator based on a silicon-LN hybrid integrated platform, with a modulation bandwidth of more than 70 GHz and a modulation rate of 112 Gbit s$^{-1}$ [18]. In order to achieve high-efficiency coupling between off-chip and on-chip LNOI devices, grating couplers [19,20], tapered waveguide or tapered fiber were introduced into the system, and coupling efficiency up to 93% [21,22] were reported.

Generally, integrated optical system includes light sources, detectors and devices with transmission and control functions. However, no laser on LNOI has been reported. Some efforts have been made to study on the LNOI devices doped with rare earth ions. The optical properties of $Er^{3+}$, $Tm^{3+}$, $Yb^{3+}$ doped LNOI have been studied [23-25]. However, due to the limitation of the concentration of ion implantation or the low quality of the integrated devices, a laser on LNOI chip is not realized.

In this paper, we fabricated on-chip erbium-doped LN microdisk cavities with high $Q$ factors ($1.26 \times 10^6$) by using UV lithography, ICP-RIE and CMP. A 1550-nm band laser output was observed under light pump in 980-nm band. The laser threshold is as low as 400 µW and the conversion efficiency is $3.1\times 10^{-4}$ %.

*Corresponding author (email: bofang@nankai.edu.cn; e-mail: kongyf@nankai.edu.cn; zhanggq@nankai.edu.cn; jjxu@nankai.edu.cn)

## 2 Fabrication and characterization of erbium-doped LNOI microdisks

We fabricated erbium-doped microdisk cavities on an erbium-doped Z-cut LNOI wafer with an doping concentration of ~0.1 mol%. The thickness of erbium-doped LN film, silicon dioxide buffer layer and silicon substrate are 0.6 μm, 2 μm, and 500 μm, respectively. Figure 1 illustrates the fabrication process of erbium-doped LNOI microdisk cavities, which is mainly divided into eight steps. To begin with, 0.4 μm thick chromium (Cr) film was deposited on LNOI wafer by magnetron sputtering method. Next, a layer of photoresist (PR) was spin coated on the Cr film. After UV exposure and development, the circular patterns on the mask were transferred to the photoresist layer. In the following, the Cr film without photoresist protection was etched up by ICP-RIE. Thus, the patterns were transferred to the Cr layer. Then, using Cr as a hard mask, the circular erbium-doped LN disks were formed by CMP. In the CMP process, a standard wafer polishing machine, polishing suspensions with silicon dioxide grains, and a very soft velvet polishing cloth were utilized. Both the Cr mask and the exposed LN film contacted with the polishing slurry due to the usage of the soft polishing cloth and the hundreds of nanometers height difference between the Cr film and LN layer. It is known that Cr (Mohs 9) is much harder than LN (Mohs 5), so the removal rate of the LN film is faster than that of the Cr film. Naturally, the circular patterns on the Cr layer were transferred to the erbium-doped LN layer. Meanwhile, the sidewalls of the erbium-doped LNOI microdisk cavities were sufficiently polished to obtain smooth sidewalls and to reduce scattering loss. Finally, the fabricated sample was immersed in a Cr etching solution for 10 min to remove the remaining Cr layer. Subsequently, the silica was partially etching in a buffered hydrofluoric acid solution to achieve suspended erbium-doped LNOI microdisks for the convenience of coupling via tapered fiber.

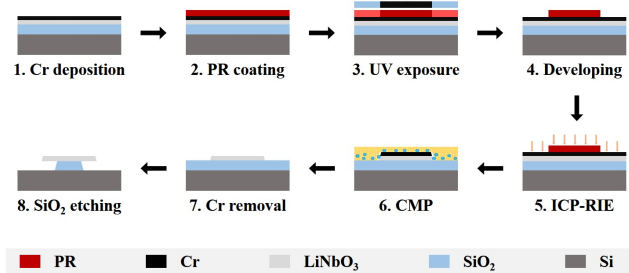

Figure 1    Schematic diagram of the fabrication process of erbium-doped LNOI microdisks. PR, photoresist; CMP, chemo-mechanical polishing; ICP-RIE, inductively coupled plasma reactive ion etching.

The geometric and optical features of the fabricated erbium-doped LNOI microdisks were characterized firstly. Figure 2(a) shows a typical optical micrograph of an erbium-doped LNOI microdisk cavity with a radius of 45 μm. Compared with undoped LNOI microdisks, we noticed that the silica pillar has a non-circular shape. This is mainly due to the loose bonding between the silicon dioxide layer and the LN film layer. As the etching time of buffered hydrofluoric acid increases, the silicon dioxide layer is eroded seriously, which may introduce additional losses to the erbium-doped LNOI microdisk cavity and reduce the $Q$ factor. Then the $Q$ factors of the erbium-doped LNOI microdisk cavities in the 980-nm band were measured by fitting the transmission spectrum of fiber-coupled microdisks. Figure 2(b) shows a typical transmission spectrum near 971.30 nm detected by using the pump-wavelength scanning method. The black curves in the figure are the experimental data, and the green, red and blue curves are fitted by Lorentzian function with multi peaks. It is shown that the loaded $Q$ value is $1.25\times10^6$ for the under-coupled mode. In under-coupling regime, the intrinsic $Q$ factor is approximately equal to the loaded $Q$ factor while the coupling loss can be neglected. We noticed that the transmission spectrum in Figure 2(b) has double peaks probably being the manifestation of mode splitting effect caused by particle scattering. The possible reason for the low $Q$ value is the relatively strong absorption of erbium ions in 980-nm band, which increases the intrinsic loss of a LN microcavity. Based on the measured $Q$ factor of the erbium-doped LNOI microdisks, we estimated the erbium concentration to be $1.9\times10^{25}$ ions·m$^{-3}$ in the case of weak pump [26]. This is close to the doping solubility of 0.1 mol% ($1.5\times10^{25}$ ions·m$^{-3}$) of the erbium-doped LN crystal from which the LNOI film was sliced.

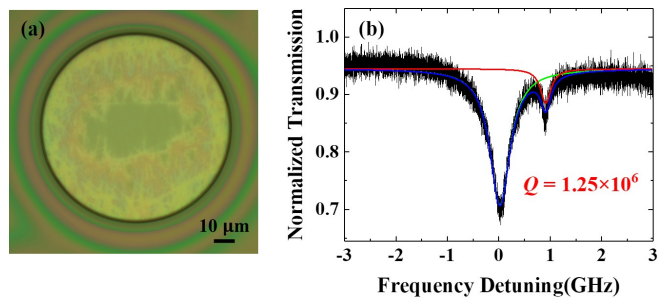

Figure 2    (a) The optical micrograph of a typical erbium-doped lithium niobate microdisk cavity; (b) The transmission spectrum showing the measured $Q$ factors of the erbium-doped lithium niobate microdisk.

## 3 Laser in erbium-doped LNOI microdisks

To investigate the photoluminescence characteristics of the fabricated erbium-doped LNOI microdisks, we pumped them with a tunable narrow-band laser operating at 980-nm band. Figure 3 schematically illustrates the experimental setup. The pump laser first passes through an optical attenuator and a polarization controller, and then is divided into two parts by using the fiber coupler 1. The small part (1%) is connected to a power meter to monitor the pump power in the optical path. The major part (99%), acting as the pump in the photoluminescence experiments, was evanescently coupled into the erbium-doped LNOI microdisk cavity through a tapered fiber with a waist of about 1 μm in diameter. The erbium-doped LNOI microdisk was placed on a three-axis piezo stage to precisely control relative position of microdisk and the tapered fiber and thus their coupling. During the coupling process, the fiber was always attached to the microdisk to ensure the signal stability. Both the generated light signal and transmitted pump were extracted by the same tapered fiber, simultaneously. Similarly, through fiber coupler 2, the output from the tapered fiber was also divided into two parts: 99% of the collected light was launched to an optical spectrum analyzer (OSA), which has a response wavelength ranging from 600 nm to 1700 nm, to detect 1550-nm band signal; 1% of the collected light was sent to a photodetector, whose electrical signal was collected and shown on an oscilloscope to monitor the transmission spectrum of the pump and thus its coupling state. In addition, through the external driving function, the sawtooth waveform voltage signal generated by the arbitrary function generator (AFG), the pump laser wavelength can be finely tuned. At the same time, the AFG provides a trigger signal to the oscilloscope, so that the transmission spectrum on the oscilloscope can be displayed stably.

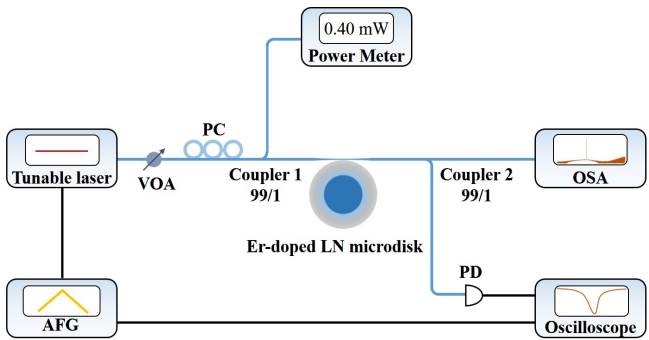

Figure 3. Experimental setup for the observation of photoluminescence in erbium-doped LNOI microdisk cavities. AFG, arbitrary function generator; VOA, variable optical attenuator; PC, polarization controller; OSA, optical spectrum analyzer; PD, photodetector.

In our experiments, light signals at the 1550-nm band were observed from OSA when the pump laser was scanned from 965 nm to 980 nm in wavelength. We also noticed that the central wavelength of the 1532-nm signal laser mode could be continuously adjusted in a range of several GHz due to the thermal broadening of pump mode [27]. The highest signal appeared around 1531.6 nm wavelength. As the input pump power increases, multiple signal peaks appear on both sides of the main signal, as shown in Figure 4(a). This is due to the fact that the metastable state $^4I_{13/2}$ and ground state $^4I_{15/2}$ of erbium ions are multiple degenerate states, which allows multiple signal peaks corresponding to energy levels in the 1550-nm band region, as shown in the right inset of Figure 4(a). We choose to detect the signal intensity at 1531.6 nm with the variation of the pump power. The central wavelength of the pump was set at 974.22 nm. Through external driving, the AFG applies a 10-mHz, 2-V peak-peak sawtooth voltage to the pump laser to finely tune its wavelength. The corresponding scanning period is 100 s, and the scanning range is from 974.07 nm to 974.38 nm, so as to adjust the pump detuning with respect to the cavity resonance to obtain the maximum signal. The signal power under different pump power were detected and shown in Figure 4(b), where the transmission loss of the components in the optical path was deducted. Obviously, the variation of the signal power with the pump power has a threshold, which confirms that the signal we observed is a laser rather than a spontaneous emission when the LNOI disk was pumped strongly. By linearly fitting the curve, we can get a laser threshold of 400 μW, and a differential conversion efficiency of $3.1\times10^{-4}$ %. In principle, the signal linewidth should be consistent with that of the cavity mode (~0.001 nm). However, due to the limited resolution of OSA (0.01 nm), the actual linewidth of the signal was not perfectly reflected.

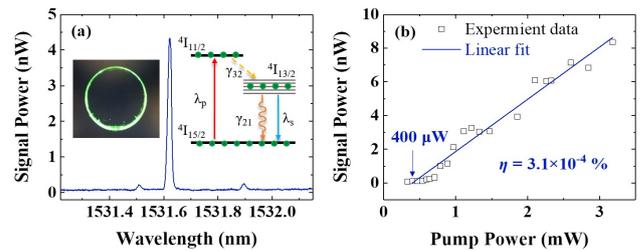

Figure 4 (a) Multi-peak signal at 1550-nm band from an erbium-doped LNOI disk under the pump of a continuous laser in 980-nm band. The left inset represents the camera image showing the up-conversion fluorescence in green. The right inset shows three-level system of erbium ions under the pump of a 980-nm laser; (b) Pump-signal power curve showing the threshold and conversion efficiency of a 1531.6-nm laser based on an erbium-doped LNOI disk.

During the experiment, we found that the appearance of the signal was accompanied by the generation of obvious green up-conversion fluorescence, as shown in the left inset of Figure 4(a). Such an up-conversion process consumes a part of pump energy, so that the gain obtained by the signal in the 1550-nm band is reduced, thereby increasing the threshold of the signal laser. In addition, LNOI film undergoes a stronger photorefractive effect under the 980-nm pump [28], creating a refractive index nonuniformity. This in turn causes the pump mode to evolve to a higher order mode, reducing the efficiency of the pump. As a result, the gain value of the signal is also reduced. In the later stage, MgO can be co-doped into erbium-doped LNOI film to suppress the photorefractive effect to further reduce the signal laser threshold.

In the preparation of this paper, we found that two similar articles were posted on Arxiv [29,30].

## 4 Conclusion

In summary, on-chip erbium-doped lithium niobate high-$Q$ microdisk cavities were fabricated. Under the pump of a continuous laser in 980-nm band, the laser operation at 1550-nm band is realized. The laser threshold is about 400 µW and the differential efficiency is $3.1\times10^{-4}$ %. This work lays a key foundation for the LNOI integrated light source and may expand the research scope based on LNOI devices.

*This work was supported by National Key Research and Development Program of China (2019YFA0705000); National Natural Science Foundation of China (12034010, 11734009, 11674181, 11674184, 11774182); Higher Education Discipline Innovation Project (B07013); The National Science Fund for Talent Training in the Basic Sciences (J1103208); PCSIRT (IRT_13R29).*